\documentclass[11pt]{article}

\usepackage[a4paper,margin=1.3in]{geometry}
\usepackage[T1]{fontenc}
\usepackage[utf8]{inputenc}
\usepackage{amsmath}
\usepackage{amssymb}
\usepackage{newtxtext}
\usepackage{newtxmath}
\usepackage[scaled=0.94,varqu,varl]{inconsolata}
\usepackage{microtype}
\usepackage{graphicx}
\usepackage{booktabs}
\usepackage[font=small,labelformat=empty,justification=raggedright,
            singlelinecheck=false]{caption}
\usepackage{needspace}
\newcommand{\vbneed}[1]{\begingroup\footnotesize\Needspace*{#1\baselineskip}\endgroup}
\usepackage{fancyvrb}
\usepackage{enumitem}
\usepackage[hyphens]{url}
\usepackage[round]{natbib}
\usepackage[colorlinks,allcolors=RoyalBlue]{hyperref}
\usepackage{orcidlink}
\usepackage[usenames,dvipsnames]{xcolor}
\usepackage{refcount}

\graphicspath{{../}}

\DeclareUnicodeCharacter{2212}{-}
\DeclareUnicodeCharacter{03C1}{\ensuremath{\rho}}
\DeclareUnicodeCharacter{03BC}{\ensuremath{\mu}}
\DeclareUnicodeCharacter{03C3}{\ensuremath{\sigma}}
\DeclareUnicodeCharacter{00D7}{\ensuremath{\times}}
\DeclareUnicodeCharacter{2248}{\ensuremath{\approx}}
\DeclareUnicodeCharacter{2264}{\ensuremath{\leq}}
\DeclareUnicodeCharacter{2265}{\ensuremath{\geq}}
\DeclareUnicodeCharacter{2192}{\ensuremath{\rightarrow}}

\fvset{fontsize=\footnotesize,xleftmargin=0pt,samepage=true}
\DefineVerbatimEnvironment{verbatim}{Verbatim}{fontsize=\footnotesize}

\setlist{itemsep=2pt,parsep=0pt,topsep=4pt}

\newcommand{\movement}[1]{%
  \vspace{1.2em}\begin{center}\normalsize\scshape #1\end{center}\vspace{0.2em}}

\title{Atom Learning Model (ALM):\\ how a real classroom got tokenised
}
\author{%
  \normalsize Philipp Bogdan\,\orcidlink{0009-0009-7405-3555}\\[3pt]
  \normalsize\itshape Department of Computing\\[1pt]
  \normalsize\itshape Imperial College London\\[1pt]
  \normalsize\texttt{philipp.bogdan25@imperial.ac.uk}%
  \thanks{The affiliation identifies the author, an undergraduate at
  Imperial College London; this report is written in a personal capacity
  and is not an output of the college. The system described was built at
  the company that operates the deployment, where the author works.}
}
\date{}

\begin{document}
\maketitle

\begin{abstract}
\noindent
The Atom Learning Model (ALM) tokenises a school curriculum. 757 pages of GCSE and
Further Mathematics material were read by machine into 1,934 atoms, each one thing a
learner can do in a single step, ordered by 4,616 machine-written prerequisite
links. Both sides of a lesson are then expressed in that one structure: a
question is a set of atoms plus everything beneath them, a child's ability is a
score between 0 and 1 on every atom of the same graph, and whether a question
suits a child is arithmetic over one index, with no difficulty parameter fitted
for either side. Nobody wrote an atom, a link or a question. Reading the 757
pages cost £55, building the whole structure cost between £615 and
£1,230, and against it the system composed 6,648 questions for
373 children in two English secondary schools over seven weeks, at 26p
per composed question. Four measurements went against expectation. The cost is
in the links, not the pages. The composer's own difficulty label has a rank
correlation of -0.0123 with measured facility, so a language model shown a
question cannot say how hard it is. Children stop working when a mark takes
seven seconds instead of three. And the deployment never served a question
deeper than two prerequisite steps, which is exactly where the central premise
becomes testable, leaving it unfalsified rather than confirmed.

\end{abstract}

\begin{center}\small
Interactive companion: \url{https://philippbogdan.com/atoms}\\[2pt]
Published data and schemas: \url{https://github.com/philippbogdan/atom-learning-model}
\end{center}

\movement{I\quad The idea}
\section{Thirty children, one hour}

A teacher standing in front of thirty fourteen-year-olds for an hour needs two
things they cannot have. They need to know, for each child, which of the
hundreds of small skills in the subject that child currently holds. And they
need to hand each child work pitched just past what they hold, which means
thirty different worksheets.

Neither is available, so the lesson is pitched at one level. English schools
narrow the gap by setting: a year group is sorted into perhaps five bands, and
each band gets work aimed at its middle. That moves the resolution from one
lesson per cohort to one lesson per thirty children, and stops there.

Content is not what is missing. Every textbook in the department is full of
easier and harder questions on every part of the syllabus. What is missing is
anything that can read a child at the resolution the questions are written at,
and anything that can produce a question aimed at one child rather than at a
band.

ALM is one answer, and it rests on a single move. The knowledge itself was
tokenised: a subject that existed as prose in textbooks and as chapter headings
in a specification became 1,934 named units with an order on them, called
atoms. A question is then written as a set of those units plus everything
underneath them, and a child's ability is written as a score on every unit of
the same graph, so the two can be read against each other directly. Sections 2
to 5 develop that idea. Sections 6 to 9 describe the build: how the catalogue
was read from 757 pages of material, how questions are composed against it, how a
mark moves a child through the graph, and what a child sees. Sections 10 to 12
describe the run: seven weeks in two schools, what it cost, what the
measurements showed, and what is not built or not measured. Section 13 is the
horizon.

The horizon is worth one paragraph up front, because it is the reason the
plumbing matters. A school that knows every child's position to the resolution
of a single step no longer needs the year group, which exists only because
thirty children must be kept together when nobody can tell where each one is.
Years become levels a child moves through at their own pace. An exam stops
being a hall booked in June and becomes something a child sits on the same
device their lessons run on, when they are ready. None of that is claimed by
the deployment described here. It is where the road goes, and section 13 states
it properly.

\section{How school mathematics is already divided}

The structure this report adds sits underneath one that every school, awarding
body and textbook already agrees on, so that one comes first.

Content is filed in a tree of five nested levels. A \textbf{syllabus} is the
qualification: GCSE Mathematics, Level 2 Further Mathematics, Key Stage 3, and
in the product this system ran inside, also English, Geography, History and
Science. A syllabus divides into \textbf{modules}, a module into \textbf{topics}, a topic
into \textbf{subtopics}, and a subtopic into \textbf{microtopics}. Figure 1 is one real
path down the GCSE tree.

\begin{figure*}[tbp]
  \centering
  \includegraphics[width=0.8\textwidth]{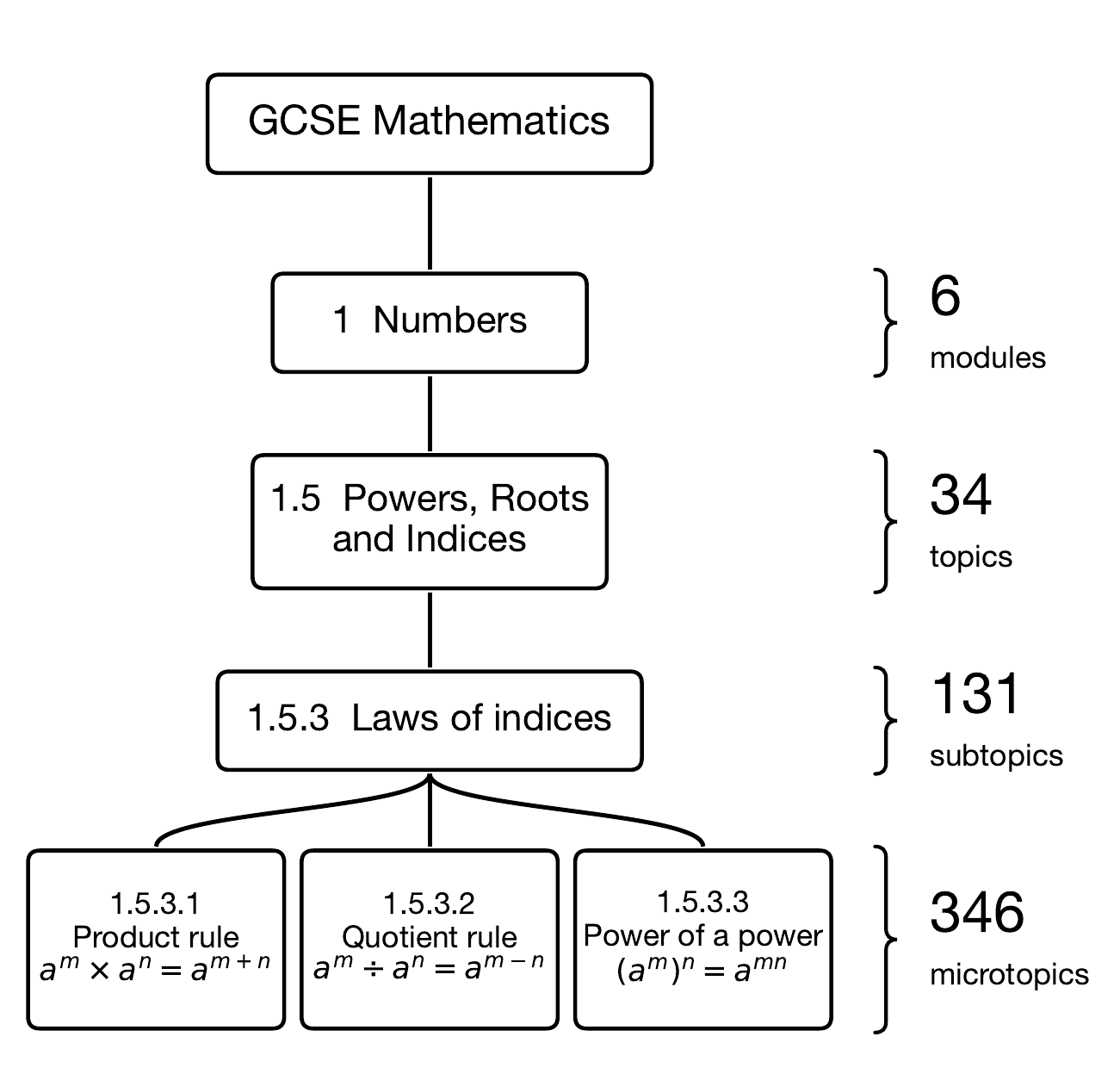}
  \caption*{\textbf{Figure 1.} \emph{One path down the tree, every label verbatim from the seed data. The counts are the whole qualification; Further Mathematics adds another 6 modules, 25 topics, 59 subtopics and 141 microtopics of the same shape. The live deployment counts 191 mathematics subtopics across the two, one having been added after these trees were seeded.}}
  \label{fig:1}
\end{figure*}

Every level of that tree is a place, not a step. It is a filing system, and a
good one: it lets a department plan a term, a publisher order its chapters, a
teacher find material, and a report tell a parent that their child is working
on ratio. What it cannot do is say what a child can do, because none of its
levels is a thing anybody does. \emph{Laws of indices} is a shelf. Several
distinguishable abilities sit on that shelf, and a child can hold three of them
and not the fourth.

That is the whole gap. A lesson is planned at subtopic resolution, which is 191
shelves across the two qualifications. A child needs work aimed at one item on
one shelf. Nothing in the tree addresses an item, so nothing built on the tree
can either.

\section{Atoms}

An \textbf{atom} is one thing a learner can do, small enough to be a single step.
The catalogue holds 1,934 of them against the 191 subtopics of the same
specification, roughly ten atoms to a shelf.

Here is one, exactly as it sits in the catalogue. It is used for the rest of
this report, so that every term lands on the same object.

\begin{verbatim}
id          ALGEBRA.SOLVE_LINEAR_SIMULTANEOUS

statement   Solve a system of two linear simultaneous equations in two
            variables algebraically using elimination or substitution,
            including multiplying equations to align coefficients.

teaching    To solve two equations, eliminate one variable. For
            elimination, make the coefficients of one variable match by
            multiplying whole equations, then add or subtract: subtract
            when the matching coefficients have the SAME sign, add when
            they have OPPOSITE signs. [...]

example     Solve 2x + 3y = 8 and 3x - y = 1  ->  x = 1, y = 2

actions     apply

home        2.4.3, 8.5.2

prereq      ALGEBRA.SOLVE_LINEAR_EQUATION
            ALGEBRA.SUBSTITUTE_AND_EVALUATE
            ALGEBRA.COLLECT_LIKE_TERMS

misconceptions
            APPLY_ADD_WHEN_SHOULD_SUBTRACT
            APPLY_MULTIPLY_ONE_TERM_ONLY
                Multiplies only one term when scaling a whole equation.
            APPLY_FORGETS_SECOND_VARIABLE
            APPLY_SIGN_ERROR_SUBTRACTING
\end{verbatim}

Four properties of that record carry the design.

\textbf{An atom is keyed to what it operates on, not to where it was found.}
Evaluating a power is one entry whether it turned up in the chapter on number,
the chapter on standard form or the chapter on growth and decay. That is what
keeps the catalogue smaller than the material it was read from, and it is why
an atom does not hang beneath one subtopic. An atom is \emph{homed} to however many
subtopics it serves; the worked atom is homed twice, to subtopic 2.4.3 of GCSE
and 8.5.2 of Further Mathematics, one skill filed in two qualifications and
paid for once. Of the atoms homed anywhere, 46.8\% serve two or more, at a
mean of 1.614 subtopics each. 325 atoms, 16.8\% of the catalogue, are homed to
no subtopic in the specification at all, and 190 of the 191 subtopics have at
least one atom under them.

\textbf{The \texttt{actions} field is what makes an atom askable.} Naming a skill does not
say what you would ask about it. Six actions appear across the catalogue:
\texttt{apply} on 1,574 atoms (do the thing), \texttt{state} on 450 (say what it is),
\texttt{identify} on 420 (recognise it in front of you), \texttt{justify} on 378 (say why it
holds), \texttt{interpret} on 199 (read what it means in context) and \texttt{derive} on 37
(get to it from something else). That gives 3,058 atom-and-action pairs, and a
pair, not an atom, is what a question is written against. Each pair carries its
own declaration of what a question on it may vary, and the ranges each varied
quantity is drawn from.

The action axis replaced an earlier taxonomy, and dissolving that taxonomy is
part of the design. The first version of the catalogue had two kinds of atom:
facts, declarative entries a learner recalls, and routines, procedures a
learner applies. The split ran deep. A worked solution carried a separate
``recall the fact'' step before every application, and marking needed per-atom
credit policies and inference rules to divide a mark between a fact and the
routine using it. The redesign collapsed both kinds into one, because the
distinction was never about the knowledge, only about how it was usually
asked: a fact was an atom mostly queried by \texttt{state} or \texttt{identify}, a routine
one mostly queried by \texttt{apply}. Once the action is explicit, the kinds carry
nothing. An atom owns its knowledge, the same atom under different actions is
genuinely different demonstrations (Pythagoras under \texttt{apply} produces a value,
under \texttt{justify} defends a claim, under \texttt{identify} spots a triple), and the
recall sub-step vanished, because the theorem stopped being a separate thing
to fetch before using it.

\textbf{\texttt{misconceptions} names the ways it goes wrong.} They were authored so a
marker could return a diagnosis rather than only a verdict. Section 12 records
that they were never used.

\textbf{\texttt{prereq} is the part that is not a taxonomy.} 4,616 links join the 1,934
atoms: 3,581 prerequisites, asserting that a learner needs the first before the
second is worth asking, and 1,035 asserting that the second is derivable from
the first. The serving policy treats the two alike. Following the worked atom's
prerequisites all the way down gives its \textbf{closure}, the set of every atom
underneath it. Expanding the bracket needs collecting like terms and the
product-of-powers law; the law needs index notation and evaluating a power;
order of operations needs fraction arithmetic. Eight atoms in all, running
three steps deep, so a question on this atom is not a flat request.

Figure 2, in the next section, draws this closure.

One defect only the whole graph can show. Deep in that closure,
\texttt{ARITH.\allowbreak{}EVALUATE\_\allowbreak{}POWER} lists \texttt{ARITH.\allowbreak{}EVALUATE\_\allowbreak{}NTH\_\allowbreak{}ROOT} as a prerequisite, and
the root atom lists the power atom back: a two-atom cycle, sitting inside the
first example this report reached for. Tarjan's algorithm, a standard linear
pass that finds every cycle in a directed graph, finds exactly three in the
3,581 prerequisite links, and all three are definitional pairs of the same
kind: power and root, comparing fractions and expressing an amount as a
fraction, defining pi and the circumference of a circle. No prompt looking at
one atom at a time can avoid these, because both directions read as true in
isolation. Only a pass over the finished graph sees them.

\textbf{Nobody wrote any of it.} Not an atom, not a statement, not a link, not a
question. People wrote four things once: the schema the model fills in when it
emits a question template, the closed list of answer forms, the small language
that constraints between a template's slots are written in, and the prompts
used to read a page. Everything else in the catalogue, and everything served
from it, is machine output.

\section{One structure, read from both sides}

This is the design decision everything else follows from, and the reason the
catalogue is worth what it costs.

\textbf{A question is expressed in atoms.} The worked atom, under the action
\texttt{apply}, plus the seven atoms in its closure. That is the whole of what the
question asks for: eight named things, in an order.

\textbf{A child's ability is expressed in the same atoms.} Mastery is that same
graph, held per child, with a score between 0 and 1 on every node. Not a score
per subtopic, not a score per topic, not a single ability number. One number
per atom per child, on the same 1,934 nodes the questions are written against.

So the two things a serving decision has to weigh live in one coordinate
system. Take a question on the worked atom and a child who holds collecting
like terms at 0.93, order of operations at 0.77 and fraction arithmetic at
0.81, but substituting and evaluating at 0.34 and solving a single linear
equation at 0.41. Whether this is the right question is no longer a judgement
call or a comparison between two invented scales. Two of the eight atoms are
missing, both are reachable, and the right move is to serve the missing atoms
first. That decision is a lookup and a threshold, per child, per question, at
no marginal cost.

\begin{figure*}[tbp]
  \centering
  \includegraphics[width=0.95\textwidth]{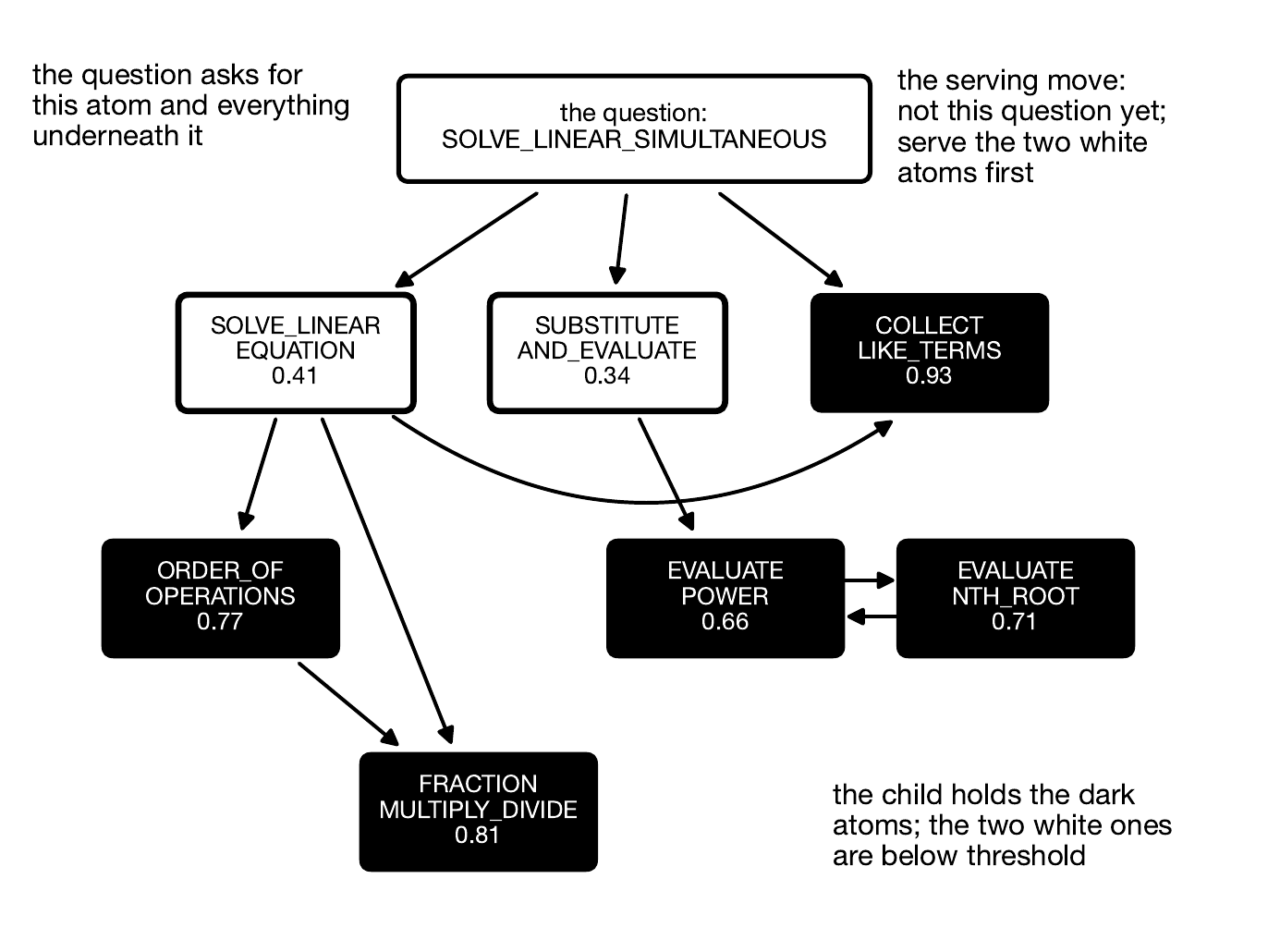}
  \caption*{\textbf{Figure 2.} \emph{The worked atom's real closure, drawn once, with one child's scores on it. The question asks for all eight; the child holds the dark six and is below threshold on the two white; the serving move follows. Scores here are illustrative, the atoms and edges are the catalogue's own, and the two-way pair at the bottom right is one of the graph's three cycles.}}
  \label{fig:2}
\end{figure*}

\textbf{What this replaces.} The standard way to decide what a learner should see is
to put the question on a difficulty scale and the learner on an ability scale,
fit both from response data, and compare the two numbers. That needs a
difficulty parameter per item, estimated from enough responses to be stable,
and it needs both scales to keep meaning the same thing across schools and
across time. ALM fits neither. There is no difficulty number on a question and
no ability number on a child. There is one structure, and two readings of it.

\textbf{The premise underneath that.} If the split into atoms is right, every atom
is about equally hard to learn on its own. What makes a question hard is not
the atom at the top but how many of the atoms underneath it the learner is
missing, and how far back the chain runs. Difficulty, on this account, is not a
property of a question at all. It is a property of the pair, this question and
this child, and it is already computable from the structure without anyone
estimating it. That premise is why the build spends 86 to 90\% of its money on
the links and nothing on difficulty labels, and section 11 is what the
deployment says about whether it holds.

\textbf{What the system actually reads.} Thirteen numbers describe where a child
stands when a question is considered, and every one is a function of the same
graph: the size of the closure under the target atom, the number of direct
prerequisites, the fraction of the closure the child is missing at each of two
thresholds, the fraction carrying any evidence at all, the mean strength across
the closure, the number of atoms the question composes, and the child's own
strength and prior attempts on the target.

\textbf{One consequence worth stating early.} Because both readings run over the
same links, a wrong link is expensive twice. It puts the wrong atoms in front
of a child, and it corrupts what the system believes the child holds, because
credit propagates down the links: a child who factorises correctly is credited
with the multiplication inside it. 62.4\% of what the system believes about
children's prerequisites arrived by that propagation rather than by direct
observation. There is no cheap substitute for getting the links right.

\section{What this is built on}

Every layer of ALM is an existing idea, and several are decades old. What is
unusual is the composition: layers that each required expert human authoring
per unit, run with none, end to end, into a live classroom. Anyone building one
should read the originals, because each solved a piece of it properly.

\textbf{A curriculum as components with an order on them.} Knowledge space theory is
the formalism \citep{doignon1985,doignon1999}, assessing against it in
practice is \citet{falmagne2006}, and two decades of ALEKS data under it is
\citet{cosyn2021}, with the structure built by an expert querying procedure
rather than by a machine reading a book. The knowledge-component framing is
\citet{koedinger2012kli}. Building the prerequisite relation by machine is a
subfield of its own: \citet{liang2015refd} read the order off the asymmetry of
Wikipedia's reference structure, \citet{roy2019prereq} infer it from unstructured
course material, and \citet{bai2025prereqsurvey} survey the decade around them. \citet{wang2016conceptmaps} extract concepts and their prerequisite map jointly from textbooks,
and \citet{huang2016textbook} build knowledge-component representations from textbook
text, which is the two halves of section 6 published a decade earlier.

\textbf{Generating questions from a specification instead of writing them.}
On-the-fly generation inside an adaptive test is \citet{bejar2002}, the
foundations are collected in \citet{irvine2002}, the item model formalised for
high-stakes use is \citet{gierl2012ijt}, and generation from a worked example
is \citet{singh2012}. Two decades of shipped mathematics template engines sit
under this in STACK \citep{sangwin2013}, WeBWorK \citep{webwork}, Numbas \citep{numbas} and Khan Academy \citep{resig2011}, \citet{greenhow2015}. In all of it a person writes the template, which is the cost language
models remove.

\textbf{Language models on the same problem.} Scoring generated items is \citet{moore2023}, tagging components onto them is \citet{moore2024}, training on
template-derived problems is \citet{zhang2024template}, generating isomorphic problems from
an existing bank is \citet{kim2025cbit}, and field-testing generated exams
at scale is \citet{isley2026}. Each starts from an item, a bank, or a
human-written component. Nearer still, \citet{wei2025kcluster} cluster a question
bank by a language-model similarity metric and \citet{duan2025kcgen} generate and
tag components for open-ended programming problems; both report machine-derived
components beating expert ones on prediction, and both start from a bank of
existing questions and induce components out of it. ALM runs in the other
direction: atoms are read out of a curriculum text before any question exists.

\textbf{Keeping the model away from arithmetic.} Changing only the numbers in a
mathematics problem moves a frontier model's accuracy \citep{mirzadeh2025}.
That result is why the composer in section 7 emits a template and a
deterministic kernel draws the values.

\textbf{A tutor holding several pieces at once.} The AC\&NL Tutor extracts knowledge
from text, builds concept maps, generates questions against them and drives an
adaptive tutor, with a person kept in the extraction loop \citep{grubisic2023acnl}. It is the nearest whole system to this one.

Every source named here was opened and checked against the claim attached to
it. Five are verified against a publisher abstract only, all pre-2013 and
paywalled: \citet{embretson1998}, \citet{gierl2012ijt}, \citet{leighton2004},
\citet{sinharay2008} and \citet{cen2006}. None carries a number quoted
in this report.

\movement{II\quad The build}
\section{757 pages into a catalogue}

Every cost in this report is in pounds, converted from the dollar billing at
\$1.30 to £1, the approximate rate across the deployment window; the dollar
originals are in the numbers ledger. The catalogue was read one page at a
time by a vision model: 757 pages of material, the GCSE course first, then
Further Mathematics. £55 in all, which is 7.2p a page; run one page
after another with no parallelism, the whole read is 5.1 hours.

\subsection{Six agents, six narrow jobs}

The build is six agents, each with one small input, one structured output, and
a contract it can be tested against alone. None of them sees the whole
catalogue at once.

\textbf{atomExtractor} reads pages and emits the atomic skills taught on them into a
single global pool. Its whole contract is what counts as one atom, and most of
its prompt is rejections: a composite (``solve a vector word problem'', which is
four atoms) is rejected, a topic label (``vectors'') is rejected, a presentation
choice (``sort numbers into a Venn diagram'') is rejected as packaging rather
than cognition, and tool use (``use a calculator to evaluate this'') is rejected
as key pressing rather than mathematics. Every identifier is prefixed by the
substrate the atom operates on, from a closed list (\texttt{ARITH}, \texttt{ALGEBRA},
\texttt{NOTATION}, \texttt{VECTOR}, \texttt{GRAPH} and the rest), never by the chapter it appeared
in. That prefix rule is what makes the pooling in section 3 happen at all.

\textbf{nearDuplicateGate} stands in front of every write. A candidate is compared
to the pool by embedding distance, and a judge can override that distance in
either direction. It fails open: a candidate the gate cannot decide on is
admitted. Deduplicating at write time is itself a lesson that cost something.
Cleaning up offline was tried on an early catalogue and did the opposite of
what it was for, re-splitting shared atoms into per-topic copies and dissolving
exactly the pooling the catalogue exists to provide. A catalogue's identifiers
are its interface, and by the time anyone gets round to cleaning them,
something has joined against them.

\textbf{subtopicMapper} joins the pool back onto the tree of section 2, one subtopic
at a time, labelling each in-scope atom with a role: \emph{focal} (newly introduced
here), \emph{sibling} (focal in another microtopic of the same subtopic, fair game
for a richer question), or \emph{prerequisite} (required, but home to a different
family, so the generator is told the chain exists and told not to write
questions on it). The union of focal and sibling across a subtopic's
microtopics is what homes an atom, which is how the worked atom ends up with
two homes.

\textbf{prereqLinker} takes one atom against the pool and emits the two relations of
section 3, deliberately kept disjoint. Conceptual prerequisites answer a
teaching-order question: what must already be known before this can be taught.
Derivable-from answers a decomposition question: if the learner knows the
constituents but never met this atom by name, could they get there by applying
them in sequence. Vector addition has arithmetic addition as a conceptual
prerequisite and nothing it is derivable from, because it is one move. Vector
midpoint is derivable from vector addition and scalar multiplication.
Collapsing the two lists is the failure this agent's prompt is written against.

\textbf{radicalProposer} makes an atom askable, one atom-and-action pair at a time.
It states the constraint the action imposes on its inputs, names the axes a
question on that pair may vary, and proposes the range each axis is drawn from.
Its output is enforced by a schema contract with reject and retry rather than
trusted. Its governing rule is that ranges belong to the pair and are never
inherited: whatever worked for the same atom's other action, or for a sibling
atom, is re-derived from this action's own constraint every time.

\textbf{teachingBackfill} takes an atom's statement, its actions and the statements
of its prerequisites, and writes the two coaching fields: the explanation a
learner could re-learn from, and the misconception bank. The bar is that a
misconception has to be concrete enough to recognise in real written working.

\begin{figure*}[tbp]
  \centering
  \includegraphics[width=0.97\textwidth]{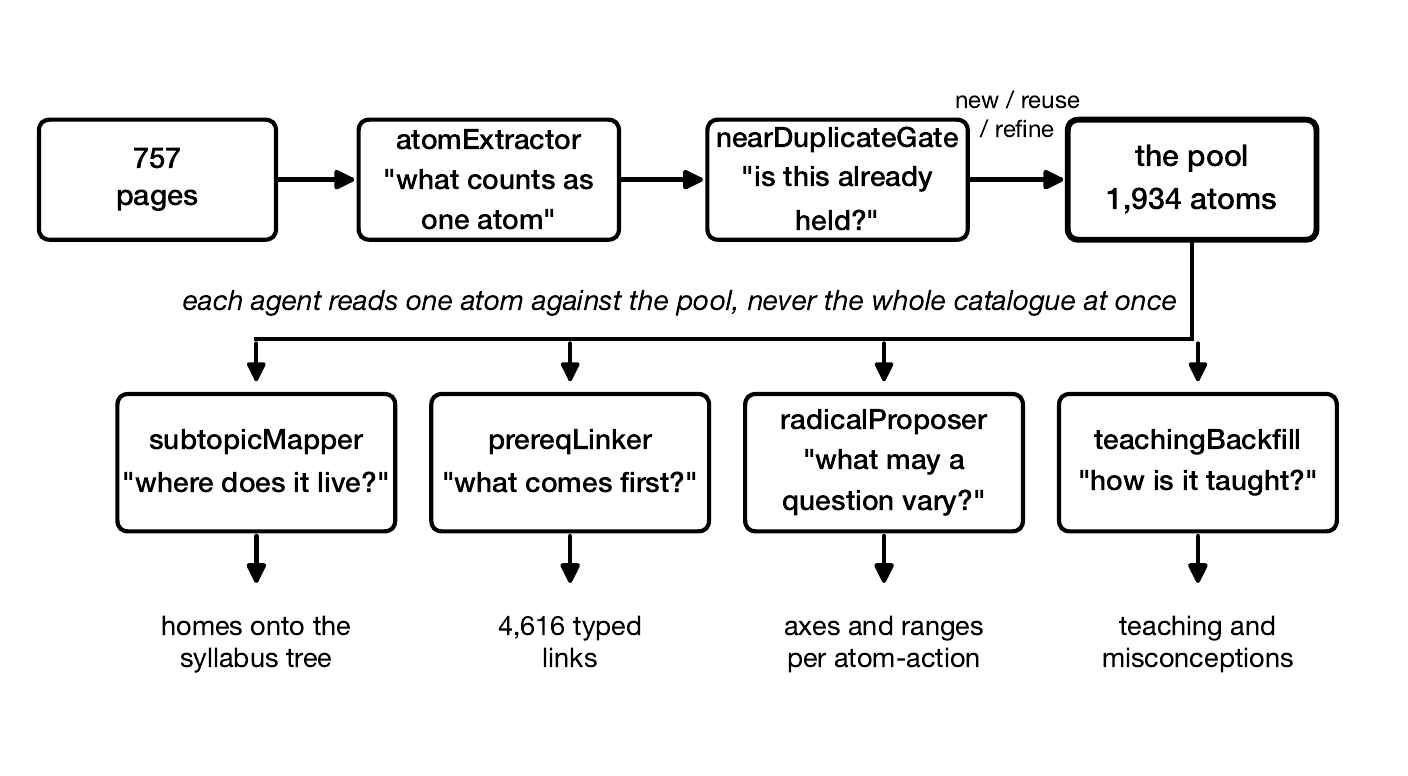}
  \caption*{\textbf{Figure 3.} \emph{The six agents. Pages flow through the extractor and the gate into the pool; the pool feeds the four agents that enrich it, each reading one atom at a time.}}
  \label{fig:3}
\end{figure*}

\subsection{What reading looks like}

A page does not produce atoms so much as opinions about atoms. Whatever the
model extracts from a page is judged against everything extracted before it,
and comes back as one of three verdicts: this is new, this is one already held,
or this is one already held but stated better. Across both courses that is 5,193
verdicts producing 1,838 new, 3,095 reuse and 260 refine.

Reuse being the majority is the payoff for keying atoms to the operation rather
than the chapter: a course says the same thing many times in different
contexts, and a catalogue that notices pays for each thing once. The pool also
does not only grow. On 35 of the 757 pages it shrank, because a refine verdict
can absorb several existing entries into one better-stated entry, and on 180
pages something already held was restated.

How repeatable is the reading? Before the production run, the same first 35
pages were read twice from the same empty pool, minutes apart. One run ended
with 85 atoms and the other with 96, and the gap between them opened as early
as page seven (Figure 4). Extraction is stochastic page by page; what is
stable is the shape of the curve and the verdict shares.

\begin{figure*}[tbp]
  \centering
  \includegraphics[width=0.78\textwidth]{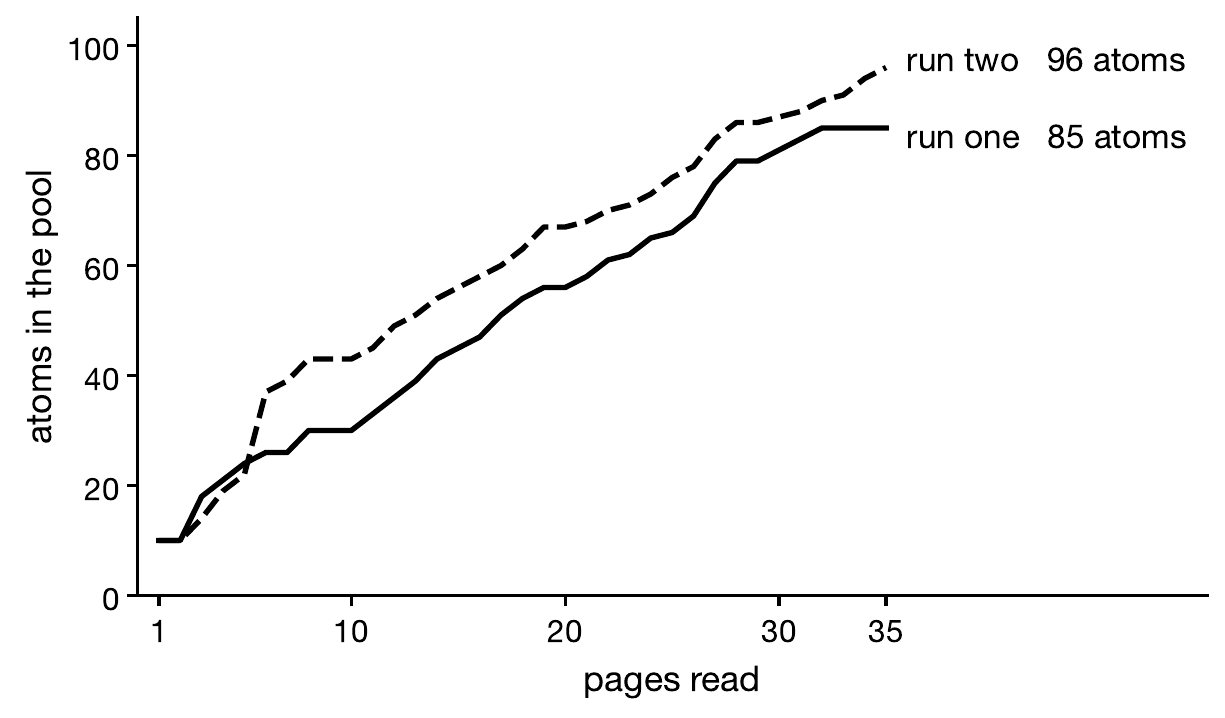}
  \caption*{\textbf{Figure 4.} \emph{The same 35 pages, read twice from the same empty pool. The runs disagree page by page and end eleven atoms apart.}}
  \label{fig:4}
\end{figure*}

The largest single event in the build sits between the two courses, and it was a
cleanup. After the GCSE material the pool stood at 2,323; an offline pass run on 28
May cut it to 1,420 before the Further Mathematics material began, removing 903
entries, which is 39\% of what the first course had built. Two things were done.
Near-duplicates were merged, and the named survivors show what the duplicates
mostly were: the extractor's own substrate-prefix rule applied inconsistently
across the first course, so the same skill entered twice under two names,
\texttt{PERCENT.\allowbreak{}REVERSE\_\allowbreak{}PERCENTAGE} folded into \texttt{ARITH.\allowbreak{}REVERSE\_\allowbreak{}PERCENTAGE},
\texttt{VECTOR.\allowbreak{}FIND\_\allowbreak{}MIDPOINT} into \texttt{GRAPH.\allowbreak{}FIND\_\allowbreak{}MIDPOINT}, \texttt{MEASURE.\allowbreak{}DRAW\_\allowbreak{}BEARING} into
\texttt{ANGLE.\allowbreak{}DRAW\_\allowbreak{}BEARING}, with the links onto the absorbed atom re-pointed at the
survivor. And over-broad entries were split into single moves: one hemisphere
atom became surface area and volume, arc length separated from the perimeter of
a sector, solving for a missing base from solving for a missing exponent.

The honest caveat is the record-keeping. Seventeen merges and twelve splits
kept named records; the other roughly 870 removals left nothing behind but the
pool count, so which atoms died, and into what, cannot be reconstructed. This
is the same lesson as the unlogged passes below, at larger scale: the
reconciliation was the single biggest edit the catalogue ever received, and it
is the least documented.

\begin{figure*}[tbp]
  \centering
  \includegraphics[width=0.85\textwidth]{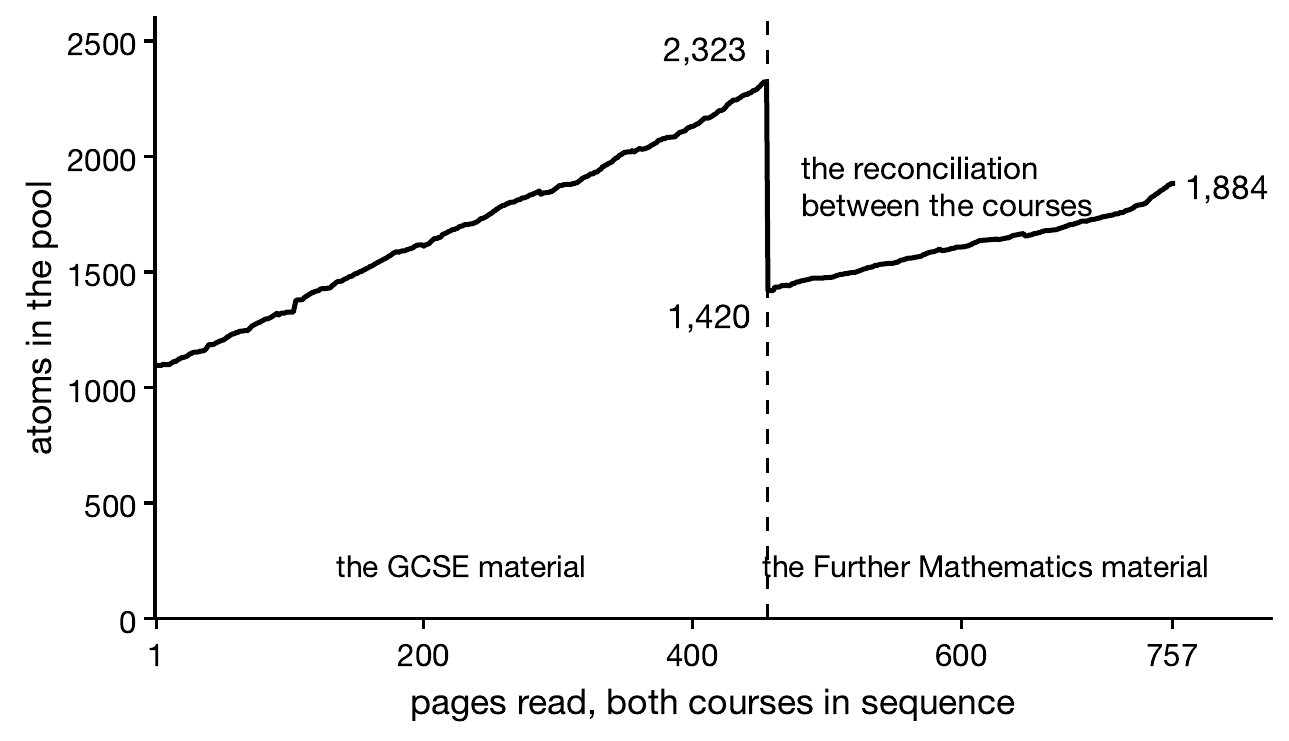}
  \caption*{\textbf{Figure 5.} \emph{The pool across all 757 pages of both courses. The line starts at 1,096 rather than zero because the GCSE run resumed on a pool built by an earlier pass whose log did not survive; the cliff at the dashed boundary is the reconciliation.}}
  \label{fig:5}
\end{figure*}

\subsection{Where the money goes}

Reading is close to flat in how much has already been read: 7.1p a page
from a starting pool of 1,096 atoms, 7.5p from a pool of 1,420, and 8.5p and
8.6p in two later gap-filling runs against a pool of 1,913. The rate rises
about a fifth as the pool grows from 1,100 to 1,900.

The links are not flat, and they are where the money goes. The two gap-filling
runs were instrumented per stage, and they partition the spend the same way:
reading the pages 8.9\% and 5.2\%, deciding what a question may vary 4.8\% and
4.6\%, wiring the prerequisite links 48.8\% and 68.3\%, joining the atoms onto the
syllabus 37.5\% and 21.9\%. The bottom two are both wiring and come to 86 to 90\%
between them. Whole-catalogue cost is therefore between £615 and £1,230 while
the reading inside it is £55: dividing £55 by reading's 5 to 9\% share
gives £615 to £1,080, and applying the gap-filling runs' all-in rate of
95p to £1.64 a page across 757 pages gives £725 to £1,240; the carried
bracket is the union of the two.

\begin{figure*}[tbp]
  \centering
  \includegraphics[width=0.9\textwidth]{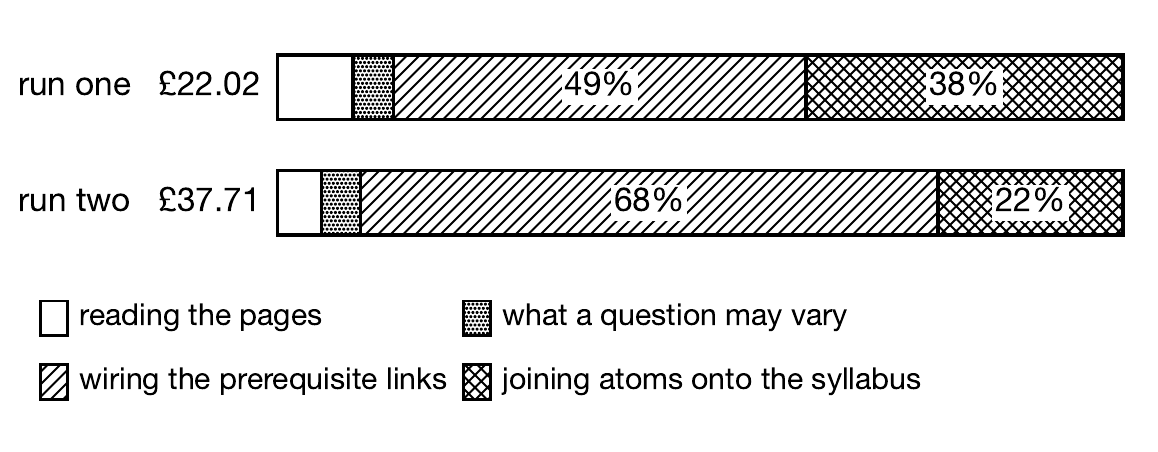}
  \caption*{\textbf{Figure 6.} \emph{Where the two instrumented runs spent, stacked to 100\%. The two wiring stages are 86 to 90\% between them.}}
  \label{fig:6}
\end{figure*}

The mechanism matters more than the split when planning a bigger build. A page
costs the same to read whether the catalogue holds a hundred atoms or two
thousand, because a page is a page. Placing a new atom costs more as the
catalogue grows, because it has to be weighed against everything already there.
Both gap-filling runs ran against a pool near 1,900, so they fix the level of
that cost and say nothing about how steeply it climbs. That slope is the number
that decides whether this is affordable across five subjects, and it is not
measured here.

One segment of the build is unlogged, and it is stated rather than folded into
the totals: the GCSE run resumed against a pool of 1,096 atoms, and no log says
what built that pool or what it cost, so £55 is what these 757 pages cost,
not what the catalogue's reading cost.

A second gap looked unlogged for months and closed under a deeper sweep of the
run directories, and the closure is worth reporting because it shows what the
snapshots buy. The ingestion runs' own counter ended at 1,884 while the catalogue
holds 1,934, and the 50 in between decompose exactly. Each run directory
snapshots the pool's identifiers at its start, and the June snapshot holds
1,913 atoms: the 1,420 that started the second course plus precisely the 493
further-mathematics atoms that survive today, so the run's counter had drifted
by 29 in its own refine bookkeeping rather than missing any real atom. A
gap-fill run on 21 June then added 21 atoms, named one by one in its log,
completing the 1,934. Its 12 June predecessor proposed 17 atoms that were
never applied; both gap-fill runs started from the byte-identical pool, and
six of the seventeen were re-found and re-added nine days later under new
identifiers. What actually remains unlogged, besides the opening pool, is
eight atoms carrying no origin stamp at all: a starter set on signed integers
and fractions that predates every logged run.

Two records were still never written and cannot be reconstructed afterwards:
which atom first appeared on which page, and what the discarded passes cost.
Log both if you build one; they are free at build time.

\section{Composing a question}

A question is built against one or more atom-and-action pairs, and the model
that builds it is not allowed to write a number.

The composer emits a \textbf{template}: question text with typed slots, a constraint
tying the slots together, an answer form drawn from a closed list, and a step
graph naming which atom is exercised at each step of the working. A
deterministic kernel then draws actual values inside the declared ranges,
checks the constraint holds, and computes the answer. The component that has to
be correct for every possible draw is the component that never sees a draw.

\subsection{The worked atom, composed for real}

What follows is a real emission for the worked atom, produced by the production
composer and replayed through the production kernel. The question text, with
its slots marked \texttt{@name@}:

\vbneed{8}
\begin{verbatim}
Solve the simultaneous equations.

    @a1@x + @b1@y = @c1@
    @a2@x - @b2@y = @c2@

Find the value of x.
\end{verbatim}

Eight slots stand behind that text, and two of them are the trick worth
learning. The coefficients \texttt{a1,\ b1,\ a2,\ b2} are drawn from 2 to 6. But the
right-hand sides \texttt{c1} and \texttt{c2} are not drawn at all: the composer declares two
hidden slots \texttt{xsol} and \texttt{ysol}, the solution itself, drawn from -6 to 8, and
derives the right-hand sides from them (\texttt{c1\ =\ a1*xsol\ +\ b1*ysol},
\texttt{c2\ =\ a2*xsol\ -\ b2*ysol}). The question is generated backwards from its own
answer, which is what guarantees every draw has a clean integer solution
without anyone checking. The constraint rules out the degenerate cases: neither
solution value zero, the two equations not parallel, coefficients not equal.
The step graph declares one step, \texttt{ALGEBRA.\allowbreak{}SOLVE\_\allowbreak{}LINEAR\_\allowbreak{}SIMULTANEOUS} under
\texttt{apply}, with the answer formula written in slot names:
\texttt{(c1*b2\ +\ c2*b1)/\allowbreak{}(a1*b2\ +\ a2*b1)}.

One real draw from the production kernel: \texttt{xsol\ =\ 7,\ ysol\ =\ 6,\ a1\ =\ 3,\ b1\ =\ 6,\ a2\ =\ 4,\ b2\ =\ 2}, giving

\vbneed{4}
\begin{verbatim}
    3x + 6y = 57
    4x - 2y = 16          answer: x = 7
\end{verbatim}

Eight seeds through the same template produced eight valid questions with
seven distinct answers. The full template, all eight draws, and the catalogue
records of the worked atom's closure are published alongside this report, so
the whole chain can be inspected end to end.

\subsection{The pipeline around it}

\textbf{The intent parser} turns a request into an envelope: which atoms, which
action, which band, whether the request is multipart, and whether it is asking
for an open response, which is guarded against separately because nothing
downstream can check one.

\textbf{The composer} emits the template above as one structured object against a
tool schema, so the output is server-validated JSON rather than text scraped
with patterns, and a malformed emission is a schema failure rather than a
parsing bug. It runs on a frontier model with extended thinking off, the
configuration its prompt was tuned under.

\textbf{Deterministic lint runs inside the composer}, with one bounded re-roll, and
some of its codes may never leave the stage: a template whose slots cancel each
other out is structurally degenerate, and the composer must produce another
rather than pass it on.

\textbf{The slot sampler} turns the template into a question. It draws candidate
values inside each declared range, sends the whole batch to the mathematics
kernel in one call, and evaluates the constraint and the answer formula for
every candidate at once. It takes the first candidate whose constraint holds
and whose answer comes out in the requested form. That reject-and-resample loop
is what produces clean numbers: the composer only has to declare honest ranges,
and the sampler finds a draw a child can reasonably be asked to work with.

\textbf{The verifier} runs its arms in parallel over the composed question: the
structural lint residue, a diagram arm, two critics (ambiguity and register), a
sanity arm, and an audit agent reading the prose against the mathematics; a
deterministic geometry check runs afterwards over the rendered diagram because
it needs the diagram arm's output. The decision rule is fixed. Residual lint, a
failing critic, or an audit verdict of wrong answer, inconsistent steps or
inconsistent guidance mean repair. A failing diagram means fail. Everything
passing means ship. Two audit verdicts deliberately do not trigger repair:
``cannot verify'' is the audit declining to decide, and churning on it wastes a
compose pass; ``error'' is infrastructure, and blocking on it would turn an
outage into a wrong-question problem.

\textbf{The repair loop} feeds the verifier's reasons back to the composer and tries
again, twice by default; a router patches known defect classes in place instead
of re-composing. When the budget runs out the pipeline reports failure and the
teacher is told the service is unavailable. It never ships the last attempt:
the alternative to a missing question is a wrong one.

\begin{figure*}[tbp]
  \centering
  \includegraphics[width=0.97\textwidth]{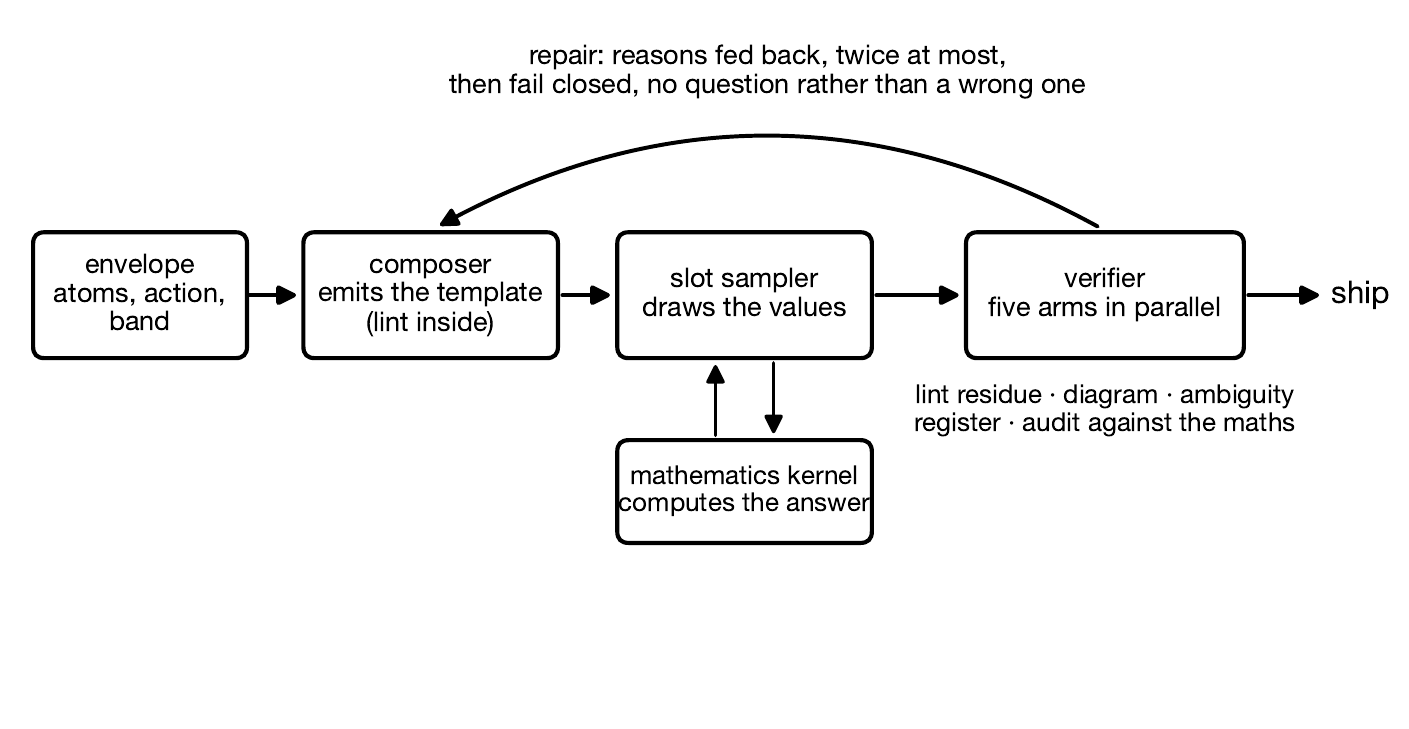}
  \caption*{\textbf{Figure 7.} \emph{The generation pipeline. The model that writes the template never sees a number; the kernel that draws the numbers never writes text; the repair loop fails closed.}}
  \label{fig:7}
\end{figure*}

Three arms were removed on evidence, and the record of what came out matters as
much as what stayed. A bespoke pedagogy critic was cut when convergence data
showed it was a net liability against the audit agent. A mark-scheme auditor
was cut after firing zero times across an ablation of ten deliberately awkward
prompts. And the arm that re-checked the answer with the kernel was cut because
it had become tautological once the slot sampler computed the answer by
evaluating the same expression: it could only fail on infrastructure flake. An
arm is a tax paid on every question forever; a contract change is paid once.
Cutting four arms alongside a tightening of the composer's contract moved the
median composition from 83.4 seconds to 26.0, the slowest question from 39.4
minutes to 6.2, compose passes per question from 1.47 to 1.16, and the
first-pass ship rate from 67.1\% to 86.5\%, while mean arms per question fell
only from 7.14 to 6.41. The arms had been absorbing defects the stricter
contract stopped producing.

\begin{figure*}[tbp]
  \centering
  \includegraphics[width=0.78\textwidth]{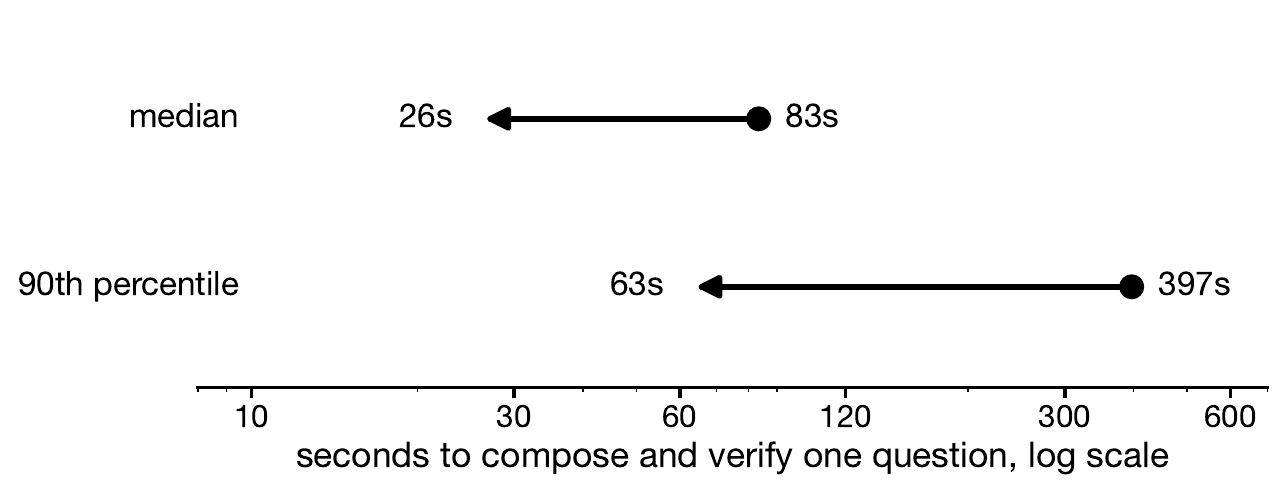}
  \caption*{\textbf{Figure 8.} \emph{What the redesign of 10 June did to composition time.}}
  \label{fig:8}
\end{figure*}

\subsection{Checking the answers, and what fails}

Every shipped question stores the formula and the drawn values it was built
from, so its answer can be recomputed from cold and compared with what the
child was shown. 6,574 of the 6,648 shipped questions can be recomputed, and
all 6,574 produce the answer that was shown. The remaining 74 cannot be
compared at all, 72 because the stored form does not support it and 2 because
the kernel errored.

A check that has never failed is worth nothing until you make it fail, so it
was made to fail on purpose: a control that corrupts every stored answer and
re-runs the comparison. It caught 6,501 of 6,574 corruptions, 98.89\%, and it
earned its keep before reporting that number, by finding four bugs in its own
comparator and one measurement reading the wrong field that had produced a
clean and entirely false result. The 53 corruptions it waved through are all
questions whose answer is a condition rather than a value, where a corrupted
value can still satisfy the condition: 36 factorisations and 17 list answers.
The rule that falls out is that an answer form must be computable from the
draw, or there is no check at all.

Composing happens in advance, never on a child's own request: every one of the
6,648 was composed into a lesson slot before a child needed it. What fails,
fails before a child sees it. The pipeline records one row per request whether
or not anything ships, and 3,172 requests survive, across 147 of the
deployment's 4,618 lessons, because the job queue ages its rows out while the
item archive is permanent. Of those requests, 79.2\% composed fresh on the
first pass, 6.4\% after a re-compose, 1.8\% were served from stock, 3.4\% fell
back to the fixed question bank the system replaced, and 9.2\% produced nothing.
Counted over questions children received, 96.3\% came from the catalogue. The
surviving lessons are the busiest ones, at 18.5 composed questions a lesson
against 1.4 across the deployment, so 9.2\% is a floor, not an average.

\begin{figure*}[tbp]
  \centering
  \includegraphics[width=0.74\textwidth]{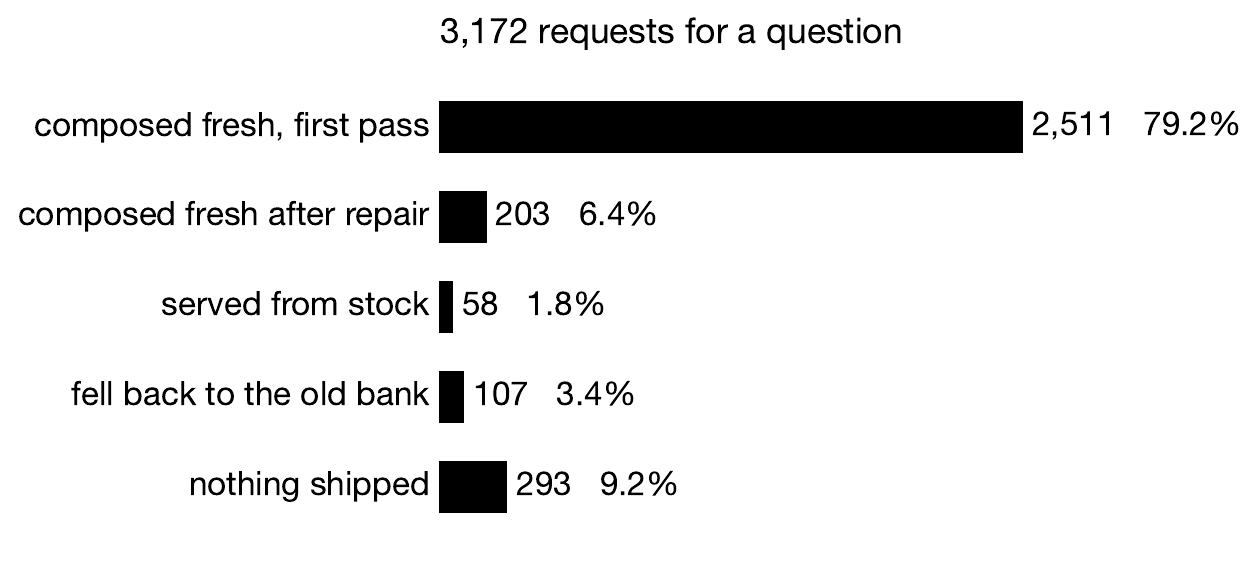}
  \caption*{\textbf{Figure 9.} \emph{Every request the queue still holds, and what happened to it.}}
  \label{fig:9}
\end{figure*}

The failure reasons matter more than the rate. Deterministic lint emits 231 of
the 455 recorded blocking reasons, led by a missing pin at 125, and read as a
bug list that is a backlog to grind down. The request shapes say otherwise:
96.6\% of the requests that died carried text a teacher had supplied, against
76.0\% of those that shipped clean, and they carried 2.77 concrete numerals
against 1.35. The three dominant codes are one collision: a request for a
template and a fixed set of values at once, which the contract cannot hold
together. On a request carrying no supplied values, none of them fires. The fix
is to detect that collision before generation starts, not to grind the 9.2\%
down afterwards.

\section{From a mark to mastery}

Marking itself is outside this report's scope. The externally built marker
reads the child's handwritten work from the tablet and returns a
mark together with per-step verdicts against the step graph the composer
declared. This section is about what the atom system does with that: how one
mark becomes movement through the graph.

\textbf{The credit vector.} Each marked attempt is decomposed into one entry per
atom the question's step graph names: the atom, the action, whether it was
credited, how strong the signal is, and which range the question drew from. The
worked question's step graph names one atom, so a correct \texttt{x\ =\ 7} produces one
credited entry for \texttt{ALGEBRA.\allowbreak{}SOLVE\_\allowbreak{}LINEAR\_\allowbreak{}SIMULTANEOUS} under \texttt{apply}. That is
typical rather than a special case: of the 16,083 credit vectors the
deployment produced, 11,006 name a single atom.

\textbf{The cascade.} Follow-through is the hard part of decomposing a mark, and it
is non-local: a child who slips early and then applies a perfect method to the
wrong number deserves credit for the method, and no verdict on one step alone
can see that. The system does not ask any model to be holistic about it. The
step graph is a directed graph with parents, so follow-through is applied
afterwards by a deterministic pass over that graph, annotating which credit is
cascading and why. Non-locality is solved by the structure, not by the marker.

\textbf{Propagation.} Credited atoms then push a share of their credit down the
prerequisite links: the child who solved the simultaneous equations is
credited, more weakly, with solving a single linear equation, substituting,
collecting like terms, and on down the closure. This is the single most
consequential mechanism in the system. 62.4\% of everything it believes about
children's prerequisites arrived by propagation rather than by direct
observation, which is what makes the mastery map dense enough to serve from,
and it is also why a wrong link corrupts belief downstream of it.

\textbf{Decay.} Per-atom mastery is a modified variant of FSRS \citep{fsrs}, the open-source
spaced-repetition scheduler: the published forgetting-curve defaults plus two
hand-set changes, a spacing weight that discounts a failure repeated in the
same sitting, and a damping of stability growth. None of the constants was
fitted, on this corpus or any other; the production configuration records them
as principled defaults awaiting a fit. A score is therefore an estimate that
falls with time and rises with evidence, not a running average of marks.

\textbf{What the deployment actually exercised.} The decomposition mechanism fired
rarely. Of the 16,083 credit vectors, 11,006 named one atom, so there was
nothing to divide. Of the 5,077 naming more than one, 2,681 credited every atom
and 1,475 refused every atom, and a vector saying the same thing about all its
atoms carries no more information than one verdict. 921 vectors, 5.7\% of all of
them, actually distinguished between the atoms of a single question. A composed
question carries 4.58 atoms on average and a credit vector carries 1.44
entries. The structure was in place; the deployment mostly served questions too
shallow to need it, which is section 12's subject.

\textbf{And the honesty that goes with the mastery table.} The per-atom scores this
section describes were computed in a four-day burst at the end of the
seven-week window, 10,591 of 11,076 rows on a single day, re-derived from the
credit vectors in each child's own time order. Nothing about the backfill is
wrong as state. What it means is that no fitted mastery ever drove what a child
was served during the deployment: the structure ran, and the scores on it were
reconstructed afterwards. Whatever decided what a child saw on 12 June, it was
not these rows.

\section{The knowledge map}

The child sees the structure directly. The knowledge map is the atom graph of
section 3 with that child's own scores on it: every atom a point, coloured by
standing, laid out so the six modules of the syllabus occupy six territories
and each module's topics cluster inside its own. Zooming resolves the next
level down, so one view runs from the whole subject to a single microtopic.
Tapping a point gives the atom's statement, its depth, its links and the size
of its closure. Holding a point draws its prerequisites in as arrows from what
comes first.

The colours are deliberately coarse: four bands, not started, just starting
from 0\%, getting there from 40\%, mastered from 80\%, because the number
underneath is a decayed estimate and not a mark. Where a child is short of
mastery, the panel converts the gap into the only unit that means anything to a
fourteen-year-old: roughly how many more questions they need to get right.

\begin{figure*}[tbp]
  \centering
  \includegraphics[width=0.85\textwidth]{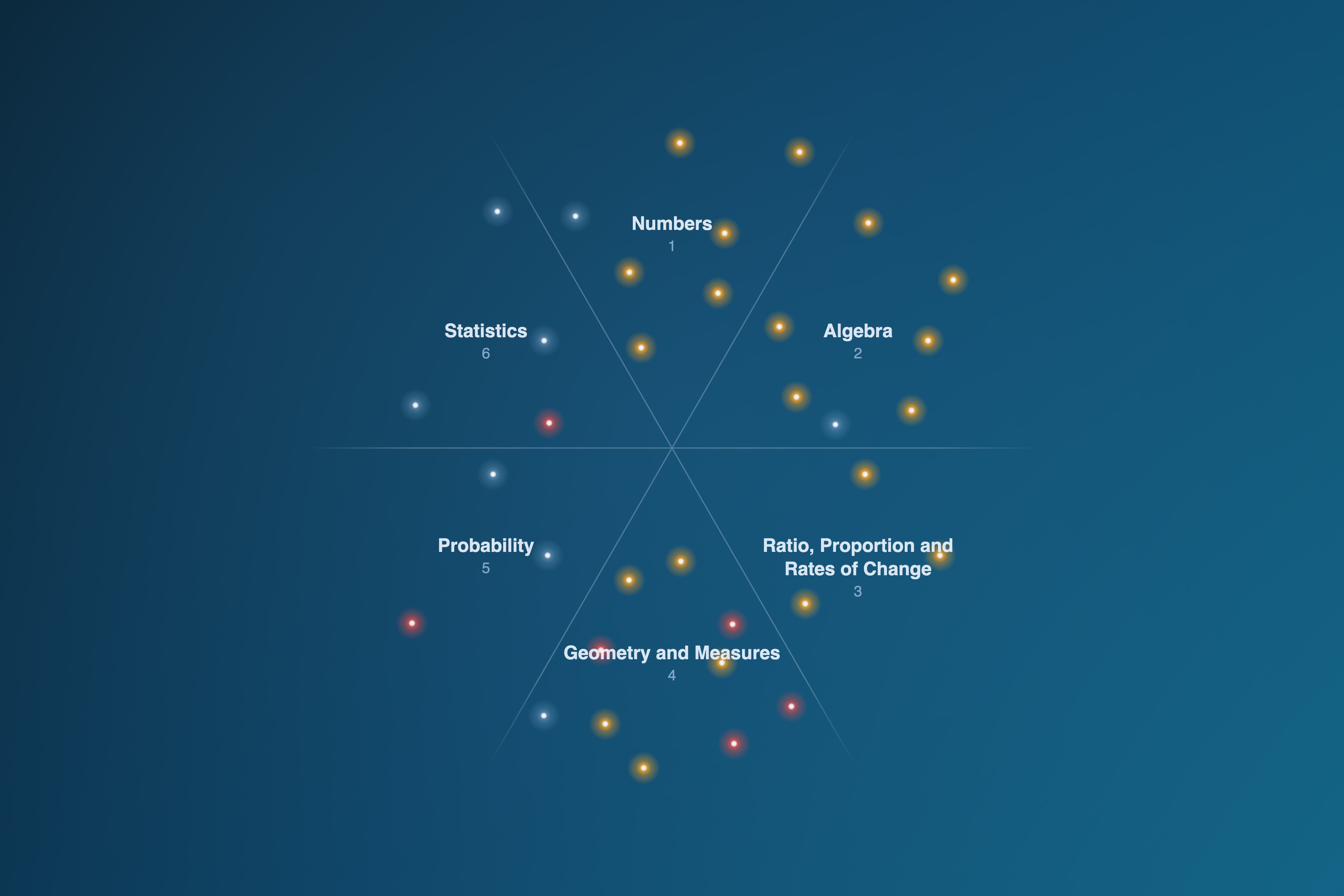}
  \caption*{\textbf{Figure 10.} \emph{The knowledge map at the whole-subject level: six module territories, their topics as points, coloured by one child's standing. Layout and colours are the app's own; the standing shown is synthetic, since the response corpus cannot leave the deployment.}}
  \label{fig:10}
\end{figure*}

\movement{III\quad The run}
\section{Seven weeks in two schools}

The system replaced something, and the changeover has a date. The same product
had been serving the same children from a fixed question bank with no atoms, no
links and no model of what a child held. The first atom was written on 27 May
2026, the first machine-composed question reached a child on 1 June, and the
last mark in the archive is 16 July. Every count in this report comes from
after the changeover.

Three counts of children recur, each a subset of the one above it, and every
claim names which one it stands on. 373 children made 26,755 attempts. 363 of
them had 26,466 attempts marked. 278 of them produced 21,761 rows, where a row
is one child's outcome on one atom of one question, so one mark makes several
rows. The 95 children in the first count and not the third are the largest
single exclusion here, and they are not a modelling choice: a row exists only
where a marked attempt left a per-atom decomposition behind. What separates
them from the 278 is volume rather than attainment, a median of 10 attempts
against 78, one active day against four; on the share of available marks
awarded they are not distinguishable, 0.640 against 0.671.

\textbf{What it cost.} £5,180 of model calls across everything the product did in
those 46 days, 311,690 calls in 74 pipeline stages. That is a ceiling on the
cost of serving questions rather than a measure of it, because the total
includes stages that belong to the wider product rather than to this system.
Composing and checking questions was £1,760 across 115,903 calls, which is 26p
and 17.4 calls per question composed and stored; widened to include the
diagram and audit stages shared with the rest of the product, 43p and 22.9
calls. The one-off build of the catalogue, £615 to £1,230, is 12 to 24\% of
the window total. Over 46 days and 373 children the whole window is £13.90 a
child, or £2.12 a child a week.

Read the per-question figure carefully, because the denominator is questions
composed, not questions answered. 26p is the cost of producing one
question including the attempts that produced nothing, which is what a team
building the pipeline pays. The archive does not record how many of the 6,648
were ever put in front of a child. The 278 children whose state reconstructs
end to end touched 1,596 distinct questions between them, and £1,760 over
1,596 is £1.10; since 1,596 is a floor on how many were answered, £1.10 is a
ceiling on what an answered one cost. The honest bracket runs from 26p to
£1.10, and a team pricing this should budget against the top of it.
Composition ran ahead of demand by design, and unanswered inventory is a real
cost line this archive cannot size any tighter.

The one-off is smaller than the running cost, and by less than it looks.
Against the line the catalogue is actually an alternative to, the £1,760 of
composing and checking, a £615 to £1,230 build is 35 to 70\%: a one-off that
pays for itself against serving somewhere between the third and the fifth week
of use. The instinct is to treat the build as the investment and the serving as
marginal, and the ratio runs the other way.

Fifteen distinct models appear across three providers, and the spend
concentrates: a cheap high-volume model does the marking (191,828 calls,
£1,810, 34.9\% of the window), a frontier model does the composing (29,871
calls, £1,320, 25.4\%), and diagrams, chat and everything else share the rest.
That split is the shape section 7 argues for: the component that must be right
for every draw is the one that runs least often. Every figure is at list price
on the day of the call, with no batching and no caching, so on cheaper tiers
expect the shape to hold and the total not to.

\section{What the measurements showed}

Six measurements the design had not predicted. One measure recurs and is worth
defining once: the \textbf{facility} of a question or an atom is the share of
attempts on it that come out correct, from 0 to 1.

\subsection{A language model cannot tell you how hard a question is}

The composer declares a difficulty band for every question it writes, and the
bands were used: the 6,648 questions split 3,378 basic, 2,002 moderate and
1,268 advanced. The design expected them to be roughly right.

They carry no information at all. Across the 730 questions with at least five
scored attempts, the \textbf{rank correlation} between the declared band and the
measured facility is -0.0123. A rank correlation asks whether two lists move
together once both are put in order: +1 means the question the model called
hardest really was answered least often, 0 means the lists are unrelated.
-0.0123 is zero.

\begin{figure*}[tbp]
  \centering
  \includegraphics[width=0.78\textwidth]{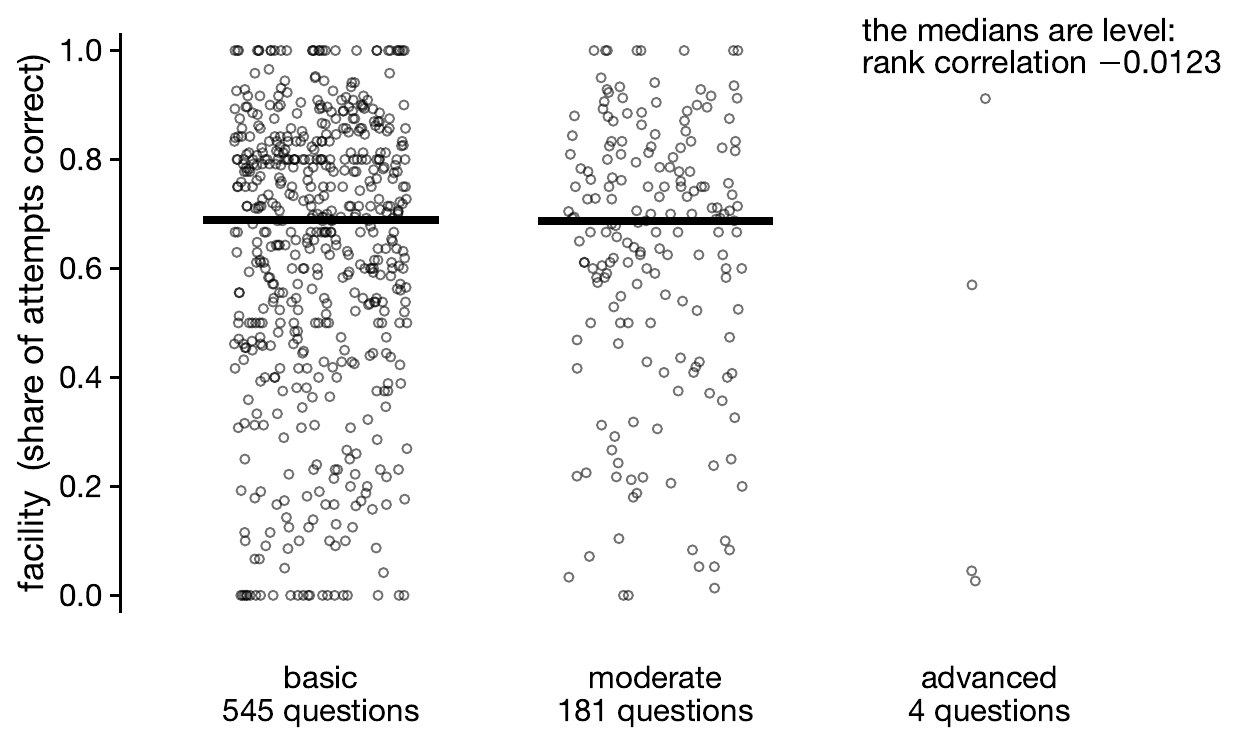}
  \caption*{\textbf{Figure 11.} \emph{Every scored question, by the band the composer declared for it. The two measurable medians are level: pooled facility 0.62 across the 545 basic questions and 0.61 across the 181 moderate ones. Only four advanced questions ever collected five scored attempts, and their pooled facility of 0.18 rests on those four, so it cannot be read beside the others. That scarcity is itself a finding: harder questions rarely got enough attempts to be measured.}}
  \label{fig:11}
\end{figure*}

That is the premise of section 4 arriving as a measurement. How hard a question
is depends on what the child in front of it holds, and a model shown only the
question is being asked for a number that is not in the question.

\subsection{The two most loaded fields were never live}

The catalogue has a field on every atom-and-action pair for a measured
difficulty, its error bar, and the number of responses behind it. 3,055 pairs
carry that field and not one was ever written. The mastery table has the
matching honesty, stated in section 8: every row backfilled after the window,
so no fitted state ever drove serving. The two halves of section 4's
unification both existed in the schema, and neither was running. What ran was
the structure without the scores on it.

\subsection{Atoms are not equally hard, and depth is worth about the same as the gap}

The premise says a well-split atom is about as hard as any other, and that
difficulty comes from depth and missing prerequisites. The deployment gives
evidence on both halves, and they pull in opposite directions.

Against the premise: atoms genuinely differ. Their facilities are spread by
about 0.23, and the spread is real rather than noise. The test is \textbf{split-half
reliability}: split each atom's responses at random into two halves, compute
the atom's facility separately in each half, and correlate the two lists across
atoms. If the differences were noise the halves would disagree; they agree at
0.877.

For the premise: one step of depth is worth about that whole spread. Comparing
questions carrying the same number of marks, facility falls from 0.638 at a
prerequisite chain depth of 2 to 0.4233 at depth 3, a drop of 0.215 against a
between-atom spread of 0.23. Depth and residual per-atom difficulty are the
same size here, and this deployment cannot separate them, because 98.1\% of the
scored questions sit at depth 1 or 2. Section 12 returns to that.

\subsection{A difficulty does not survive a move between schools}

The two schools sit on different awarding bodies. On the 58 atoms both attempt
at least fifteen times, the facility measured at one school barely predicts the
facility measured at the other. The obvious objection is that the schools have
different children, and the check for that is \textbf{differential item
functioning}: compare children matched on their own overall attainment, and
ask whether two equally strong children from the two schools still succeed at
different rates on the same atom. 18 of the 50 atoms for which that comparison
is defined function differently by a standard measure, against 4.3 expected
when the school labels are randomly shuffled. A build that stores one
difficulty per atom and serves it everywhere is storing something that did not
survive a move across two schools in the same country on the same subject.

\subsection{Four seconds of waiting decides whether a child carries on}

Marking latency moves within a lesson for reasons that have nothing to do with
the child, queue depth and network conditions among them. Sorting the 26,466
marked attempts into five equal groups by how long the child waited for their
mark, the share who started another question in the same session falls from
0.941 at a median wait of 3.2 seconds to 0.760 at 7.3 seconds, stepping down
through every group between.

\begin{figure*}[tbp]
  \centering
  \includegraphics[width=0.78\textwidth]{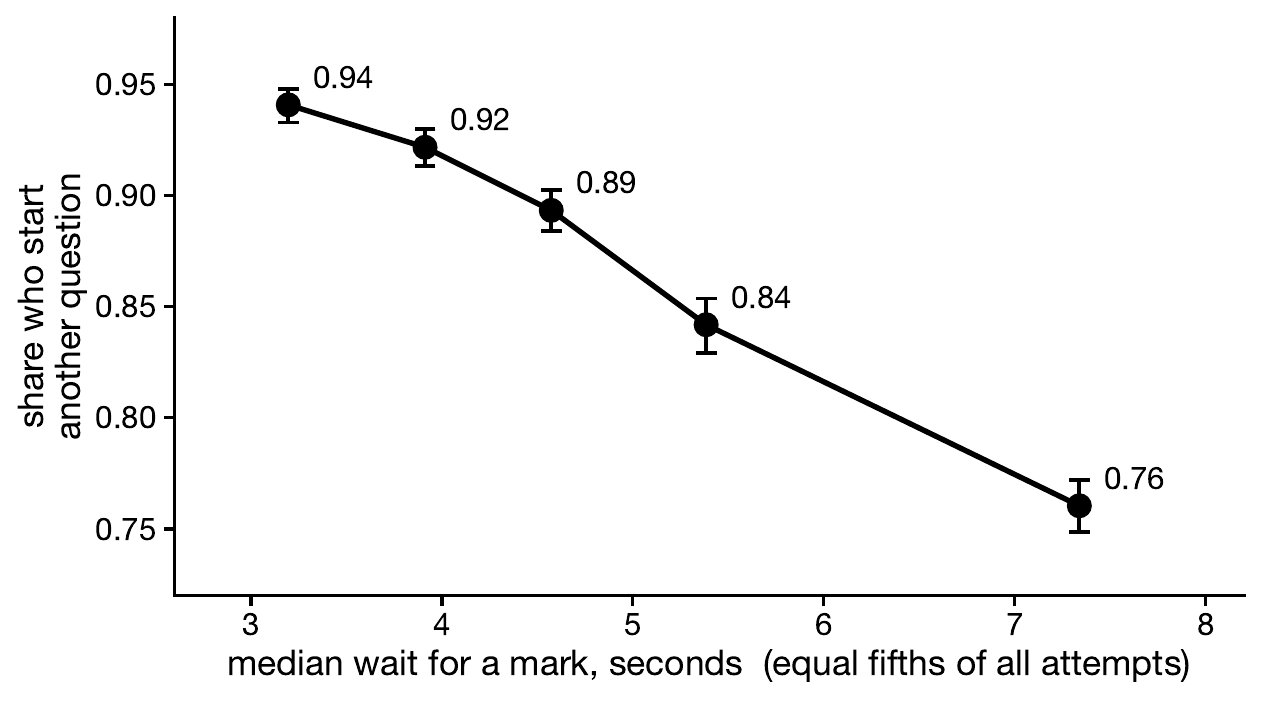}
  \caption*{\textbf{Figure 12.} \emph{The share of attempts followed by another attempt in the same session, against the wait for the mark, in equal fifths of all 26,466 marked attempts. Bars are 95\% intervals.}}
  \label{fig:12}
\end{figure*}

Holding fixed how much the child wrote, how many marks the question carried,
what they scored and how far into the session they were, doubling this
attempt's wait is associated with a fall of 13.55 \textbf{percentage points} in that
probability, where a percentage point is a hundredth of probability. The wait
for the previous attempt carries only 4.45 points, and that is the control: a
delay already sat through consumed the same lesson time and moved the child the
same distance towards the bell, so it bounds how much of this is a session
simply running out. The 9.10 point difference is what the finding rests on, and
it survives adding elapsed session time and a curve in session position.
Nothing was randomised, so this is an association, and two further results
narrow the alternatives: a longer wait does not lower the child's next score,
and after a longer wait children write more, not less. What changes is whether
they start the next question at all.

The whole effect lives between 3.2 and 7.3 seconds, inside a band most teams
would call fast and never think to defend. A specification written at the
median cannot see it. Specify the slow tenth.

\subsection{A generator scored on its own claim inflates the claim}

Under an early prompt, 7.9\% of the structural claims a generator made about its
own questions were fabricated, and the inflation rose with how much structure
was asked for. So nothing about a question's structure is taken from the
generator's word: depth is measured from the artefact, as the longest path
through the step graph, with every dependency edge re-derived by parsing which
earlier steps each expression refers to. One model generation later the
incentive response had gone, which is why the re-derivation stays and the check
repeats per generation. It is nearly free, and it is the only instrument that
sees the problem at all.

\section{Not built, not measured}

\textbf{The deployment never served a deep question.} 98.1\% of the scored questions
sit at a prerequisite chain depth of one or two; 14 questions in total reach
depth 3 or more, and the entire evidence base past depth 4 is 85 attempts from
7 children. That is exactly where the premise of section 4 would have been
settled: on a question composing six atoms four steps back, ``difficulty is the
atom'' and ``difficulty is what you are missing'' predict different things, and on
a one-step question they predict the same thing. The catalogue holds the depth
and the composer can address it; the deployment never asked for it. The central
claim is unfalsified rather than confirmed, and serving deep questions on
purpose is the first thing to do next.

\textbf{It cannot pose an open-ended question.} Every shipped question carries an
answer formula and a step graph, both fixed at composition time: the composer
decides one method, the kernel computes that method's answer, and the work is
checked against it. That is what makes the pipeline verifiable, and it is
exactly what rules out a proof, which has many valid routes and no single
answer value. What is missing is a checker that can accept a route the composer
never wrote down: several admissible step graphs at once, and a decision about
which one the child took before any decision about whether they took it
correctly. Until that exists, the system covers the part of a mathematics
curriculum that is computation, not the part that is argument.

\textbf{The state was not live.} Sections 8 and 11.2 record it: the mastery scores
were reconstructed after the window closed, so the deployment never actually
served off the structure's second reading. Making the state live is plumbing,
not research, and until it is done none of this has been tested doing the thing
it was built for.

\textbf{No qualified person has checked the catalogue.} No teacher, examiner or
subject specialist has looked at an atom, a link, or a question's alignment to
the specification and said whether it is right. Two controls say something
weaker. Replaying the seven weeks through links with the same shape, the same
number of edges and the same count at every node, but attached to the wrong
atoms, makes the system's predictions about children worse, and the real links
beat twenty such shuffles in twenty attempts; turning the downward propagation
off entirely also makes them worse. That is a \textbf{permutation test}: shuffle the
thing being tested, re-run everything, and count how often the real version
wins. Twenty shuffles floor the strongest reportable result at one in
twenty-one, and beating a shuffle is not the same as a link being pedagogically
correct, which matters most exactly where a system like this could do harm: a
wrong link puts a child in front of work they have no grounding for.

\begin{figure*}[tbp]
  \centering
  \includegraphics[width=0.85\textwidth]{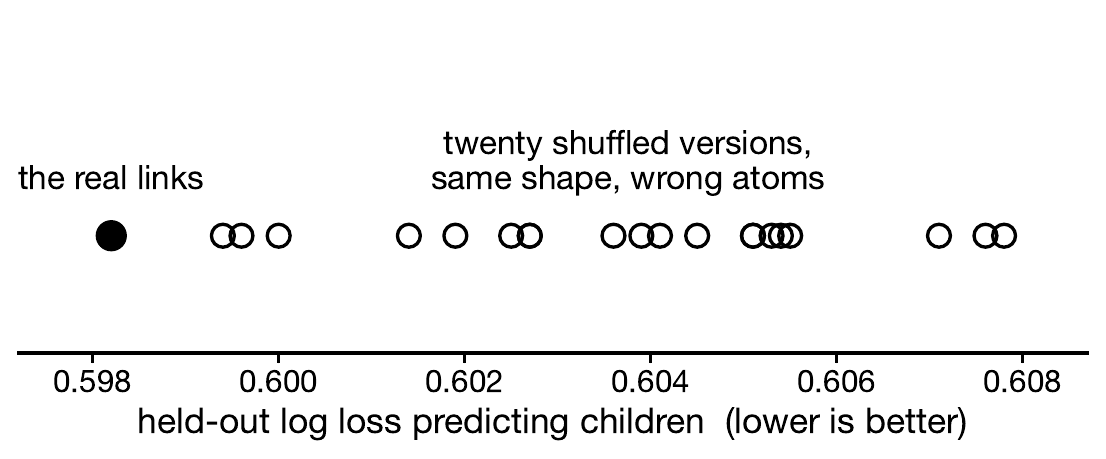}
  \caption*{\textbf{Figure 13.} \emph{The real prerequisite links against twenty shuffled versions with the same shape attached to the wrong atoms, scored on how well the model predicts held-out attempts. The real links beat every shuffle.}}
  \label{fig:13}
\end{figure*}

\textbf{The marker's accuracy is an unmeasured input.} No human re-marked a sample
of the 26,466 attempts, and since the credit vector is the only thing the model
of the child reads, every number about a child here inherits that unknown. A
few hundred hand-marked attempts would close it, and it is the cheapest missing
measurement in this report.

A team with their own deployment needs only three tables to rebuild every
number in sections 10 and 11 against their own logs: one row per attempt (child
index, atom, credited or not, when, prior attempts on that atom, how much of
the closure carried evidence at that moment); one row per question (the atoms
it composes, its declared band, compose passes, cost); one row per session
(child index, ordered attempts, the wait for each mark, whether another attempt
followed).

\movement{IV\quad The horizon}
\section{A school without year groups}

Everything above is a system that serves a better next question. The reason to
finish it is what stops being necessary once every child's position is known to
the resolution of a single step.

School years exist because a cohort has to be kept together, and a cohort has
to be kept together because there is no way to know where thirty children
individually are. Remove that constraint and the year stops being a container
and becomes a label on a level. There would still be thirteen levels of school.
A child would no longer be required to spend exactly one year on each.

In that arrangement a child works at whatever sits just past what they hold,
which is the only place learning is efficient, and moves when they are ready. A
child might cover a level in nine months and cross into the next level's
material before the summer; another might take eighteen months over the same
level and arrive knowing it properly rather than arriving on time knowing it
badly. Both are normal, and neither is a concession, because the structure
knows what each of them holds either way.

Assessment is what currently makes this impossible, because an exam is a
logistical event: thirty children, one hall, one morning, one paper. If the
exam runs on the same device the lessons run on, it stops being an event and
becomes something a child sits when they are ready, in a room that is always
available for it. No cohort has to be assembled, because no two children need
the same paper on the same day.

A class then stops being a fixed group of thirty children born in the same
twelve months and becomes something closer to a gym: a room, a teacher, and
whoever is working on that material at that time. The teacher's job gets harder
to describe and easier to do, because the thing they lost, knowledge of what
each child in front of them can do, is the thing the structure gives back.

None of this follows from seven weeks in two schools, and none of it is claimed
by them. It is what the curriculum being tokenised is for: the year group, the
set, the hall in June are all workarounds for not knowing where each child
stands, and a school that knows can stop paying for them.

\label{docend}
\typeout{ENDPAGE=\getpagerefnumber{docend}}

\bibliographystyle{acl_natbib}
\bibliography{../refs}

\end{document}